\documentclass{PoS}
\usepackage{cancel}
\usepackage{subfigure}

\title{Step Scaling with Off Shell renormalization}

\ShortTitle{Step Scaling with Off Shell renormalization}

\author{\speaker{Rudy Arthur}\\
        University of Edinburgh\\
        E-mail: \email{R.Arthur@sms.ed.ac.uk}}

\author{Peter A. Boyle\\
        University of Edinburgh\\
        E-mail: \email{paboyle@ph.ed.ac.uk}}

\abstract{A method for computing renormalization constants in the Rome Southampton scheme
with volume sources and arbitrary momenta is described. This new method is found to
enable controlled and precise continuum extrapolations and opens the way to compute the running of operators non-perturbatively in the Rome Southampton scheme. We describe this in detail and exhibit several examples of lattice step scaling functions.}

\FullConference{The XXVIII International Symposium on Lattice Field Theory, Lattice2010\\
		June 14-19, 2010\\
		Villasimius, Italy}

\begin{document}

\section{Introduction}
Regularization invariant (RI/MOM) schemes \cite{Martinelli:1994ty} are extremely useful for renormalizing matrix elements in lattice gauge theory. They are simple to implement for arbitrary operators and there are perturbative calculations, \cite{Chetyrkin:1999pq} \cite{Sturm:2009kb}, up to three loops relating RI/MOM to $\overline{MS}$. Using momentum sources, \cite{Gockeler:1998ye}, excellent statistical precision is also obtained. The leading uncertainties are systematic effects: spontaneous chiral symmetry breaking and $O(4)$ breaking. Chiral symmetry breaking leads to a $\frac{1}{p^2}$ pole, violation of Ward identities and a strong mass dependence in the vertex functions. Subtracting the condensate term is possible \cite{Noaki:2009xi}, however a change in kinematics is sufficient to suppress almost all of the chiral symmetry breaking effects \cite{Aoki:2007xm}. In \cite{Aoki:2007xm} the incoming and outgoing momenta are given the same magnitude and different directions so that the momentum transfer is non-zero and equal in magnitude to the external momenta. These non-exceptional kinematics are sufficient to suppress spontaneous chiral symmetry breaking by a factor $\frac{1}{p^6}$.

With these improvements $O(4)$ symmetry breaking is now the main source of uncertainty. We propose to use twisted boundary conditions to remove it. This has immediate impact in reducing the error on many phenomenologically important parameters where renormalization is a large and often dominant contribution to the total uncertanity. In this work we discuss the quark mass \cite{Antonio:2007pb} and kaon bag parameter, $B_K$ \cite{Antonio:2007pb, Allton:2008pn, Kelly:2009fp}, but the technique also has relevance for many RBC-UKQCD calculations, for example $K \rightarrow \pi \pi$ matrix elements \cite{Blum:2001xb}, distribution amplitudes \cite{Donnellan:2007xr, Boyle:2006pw} and nuclear form factors \cite{Aoki:2008ku}.

The largest remaining systematic error is perturbative. Usually the simulated momenta must satisfy the condition,
\begin{equation}\label{eq:window}
\Lambda_{\rm QCD}^2 \ll p^2 \ll \left(\frac{\pi}{a}\right)^2.
\end{equation}
The lower bound is for convergence of perturbation theory while the upper is in order to have small discretization artefacts. We propose to control the discretization error by continuum extrapolation, convergence of perturbation theory can then be improved by choosing a high momentum scale before matching. To increase the energy scale while avoiding large lattices we will step scale \cite{Luscher:1991wu} and we outline how to do this using RI/MOM renormalization.

\section{Twisted Boundary Conditions}
On the lattice $O(4)$ symmetry is explicitly broken to $H(4)$. This means the same quantity computed using inequivalent momentum directions will have a different Symanzik expansion depending on the direction. Matching the same physical momentum and the same direction on two different lattices with this constraint is generally difficult. In order to chose momenta with arbitrary magnitude in a given direction we use twisted boundary conditions \cite{Flynn:2005in}. Consider Green's functions of bilinear operators $\bar{q}(x) \Gamma q(x)$, $\Gamma$ is a Dirac matrix. With twisted boundary conditions the quark field satisfies,
\begin{equation}
q(x+L) = e^{i\theta}q(x).
\end{equation}
Let $\tilde{q}(x) = e^{iBx}q(x)$ with $B = \frac{\theta \pi}{L}$. This modifies the Dirac operator,
\begin{equation}\label{dirac}
D = (\cancel{\partial} + M) \rightarrow \tilde{D} = (\cancel{\partial} + i\cancel{B} + M).
\end{equation}
The inverses are related by $S(x,y) = e^{iB(x-y)} \tilde{S}(x,y)$. The momentum source method gives us the Fourier transformed propagator, $\tilde{G}(z,p)$, where $p$ is a Fourier mode and
\begin{equation}
\sum_z \tilde{D}(y,z)\tilde{G}(z,p) = e^{ipy}.
\end{equation}
$\tilde{G}(z,p)$ is related to $G(z,p)$ , the Fourier transformed propagator, by
\begin{equation}\label{Gtilde}
\tilde{G}(z,p) = \sum_x e^{-iB(z-x)} S(z,x) e^{ipx} = e^{-iBz} G(z,p+B).
\end{equation}
Thus, solving the Dirac equation with a plane wave source and twist $B$ gives the momentum space propagator $G(z,p+B)$. $B$ is arbitrary and so allows arbitrary momentum in any direction. 
 
To compute vertex functions $\Lambda_\Gamma$ we use,
\begin{equation}\label{Gp1p2defa}
  G_{\Gamma}(p_1+B_1, p_2+B_2) = \sum_{x,y} \gamma_5 e^{i (p_1+B_1)x}G(x,p_1+B_1)^\dagger \gamma_5 \Gamma e^{-i (p_2+B_2)y}G(y,p_2+B_2).
\end{equation}
\begin{equation}
\Pi_{\Gamma}(p) = \left( G^{-1}(p_1,p_1) G_{\Gamma}(p_1,p_2) \gamma_5 [G^{-1}(p_2,p_2)]^\dagger \gamma_5 \right)\end{equation}
\begin{equation}
\Lambda_{\Gamma} = \frac{1}{12} Tr \left( \Pi_{\Gamma} P_{\Gamma} \right).
\end{equation}
$P_\Gamma$ is a projector\cite{Sturm:2009kb}. We pick $p_1 = (-1,0,1,0) \, \,p_2 = (0,1,1,0)$ to minimize $\sum_i p_i^4$.

\begin{figure}[htp]
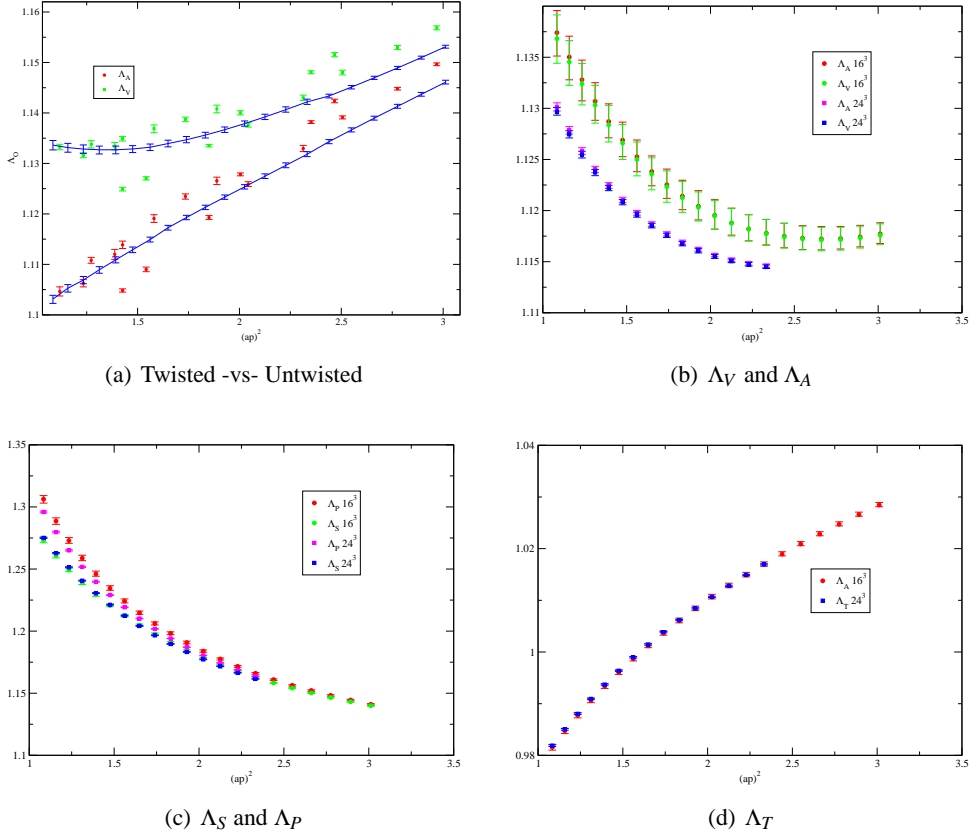

  \begin{center}
\subfigure[Twisted -vs- Untwisted]{\includegraphics[width=0.4\textwidth]{Zgf_E_0220_0.03_cf.eps}} \,\,\,\,\,\,\,\,
\subfigure[$\Lambda_V$ and $\Lambda_A$]{\includegraphics[angle=0,width=0.4\textwidth]{VAvol.eps}}\\[5mm]
\subfigure[$\Lambda_S$ and $\Lambda_P$]{\includegraphics[angle=0,width=0.4\textwidth]{SPvol.eps}} \,\,\,\,\,\,\,\,
\subfigure[$\Lambda_T$]{\includegraphics[angle=0,width=0.4\textwidth]{Tvol.eps}}

  \end{center}
  \caption{ (a) The axial(red) and vector (green) verticies computed at exceptional momentum \cite{Martinelli:1994ty} with a volume source at fixed quark mass $m_q = 0.03$ on a $16^3 \times 32 \times 16$, $\beta = 2.13$ lattice, using 10 configurations \cite{Arthur:2010ht}. The red and green points use Fourier modes while the blue data uses twisted boundary conditions to vary the momentum. Twisting removes the $O(4)$ breaking scatter. (b,c,d) The bilinear vertex functions in the chiral limit computed on $16^3$ and $24^3$ lattices. All parameters except volume are identical. The volume dependence is strongest for the axial and vector currents although it is still only a fraction of a percent}
  \label{fig:voldep}
\end{figure}

Figure \ref{fig:voldep} (a) shows the effect of the twisted boundary conditions on the data. The $O(4)$ breaking scatter is removed. We simulate momenta with one orientation and continuously varied magnitude so that the vertex function is a smooth function of $p^2$. The fact that we have a valid Symanzik expansion then enables us to take the continuum limit in an unambiguous way by choosing the same direction on every lattice.

\section{Step Scaling with RI/MOM}

Twisted boundary conditions mean that we can now study the same physical quantity, at the same scale with the same Symanzik expansion on any lattice. This enables a controlled continuum extrapolation of renormalized quantities $Z_O \langle O \rangle$. Using continuum extrapolation to control discretization effects the majority of the remaining uncertainity comes from perturbation theory.

Rather than try to satisfy the window condition \ref{eq:window} on a single lattice we consider a series of lattices of successively descreasing volume and use these to step up to high energy. On each lattice we still need to compute vertex functions with small discretization error. The lower limit on the momenta is now given by the requirement that we do not resolve the finite volume of the lattice. This gives the step-scaling window condition,
\begin{equation}
\label{eq:sswindow}
(\frac{\pi}{aL})^2 \ll p^2 \ll \left(\frac{\pi}{a}\right)^2
\end{equation}
By simulating a series of lattices with overlapping scaling windows we will be able to compute continuum limit step scaling functions. 

Explicitly, on each lattice at a given quark mass we can compute the ratio,
\begin{equation}
R_{O}(p,a,m) = \frac{\Lambda_A(p,a,m)}{\Lambda_{O}(p,a,m)} = \frac{Z_{O}(p,a,m)}{Z_A(p,a,m)}.
\end{equation}
here $\Lambda_A$ is the axial vector vertex function. We use it to divide out a factor of the field renormalization $Z_q$ in favour of $Z_A$ which can be computed more accurately. The chiral limit is
\begin{equation}
Z_{O}(a,p) = \lim_{m\to 0} Z_A(p,a,m) R_{O}(p,a,m)
\end{equation}

All of the scale dependence is in the renormalization constant and so the factor needed to change the scale from $p$ to $sp$, where $s$ is a constant, is
\begin{equation}
\Sigma_{O}(p,s p,a) = \lim_{m\to 0}
                        \frac{R_{\cal O}(sp,a,m)}{
                             R_{\cal O}(p,a,m)}  
			= \frac{Z_{O}(a,sp)}{Z_{O}(a,p)}.
\end{equation} 
Continuum extrapolating this gives
\begin{equation}\label{eq:sig}
\sigma_{O}(p,s p) = \lim_{a\to 0}\Sigma_{O}(p,s p) = \frac{Z_{O}(sp)}{Z_{O}(p)} = exp\left( \int_{\alpha(p)}^{\alpha(sp)} \frac{\gamma(x)}{\beta(x)} dx \right).
\end{equation}
A series of $n$ lattices to each raise the scale by a factor $s$ gives
\begin{equation}
\begin{array}{ccc}
\langle {O}^{\overline{MS}}(\mu) \rangle
&=&
\langle {O}^{\rm SMOM}(p) \rangle \times \sigma^1_{O}(p,s p) \ldots \times \sigma^n_{O}(s^{n-1}p,s^np)\\
&\times&
\left[1 + c^{\rm SMOM\to\overline{MS}}\alpha_s(\mu = s^n p)\right].
\end{array}
\end{equation}
Hopefully, perturbation theory is well convergent at the last step where the scale is large.

\subsection{Setting the Scale}

In order to set the lattice scale accurately a different definition of lattice spacing for each volume is needed. This means that between two successive steps the same momentum $s^n p$ will be defined in different ways. The two definitions must be such that they agree in the continuum limit so that products like $\sigma^n_{O}(s^{n-1}p,s^np) \sigma^{n+1}_{O}(s^{n}p,s^{n+1}p)$ are well defined. We need a family of scale setting quantities $\{ q_i(a) \}$ that depend on shorter distances as we reduce the volume between steps.
\begin{equation}
\frac{s^n p}{q_n(a)} =  \frac{s^n p}{q_{n-1}^{\rm cont}}  \left( \frac{q_{n-1}}{q_n} \right)^{\rm cont} , 
\left( \frac{q_{n-1}}{q_{n}}\right)^{\rm cont} =  \lim_{a\to 0} \left( \frac{q_{n-1}(a)}{q_{n}(a)}\right)^{\rm cont}
\end{equation}
we ensure that scales set on lattice $n$ using $q_n$ agree in the continuum with scales set using $q_{n-1}$.

We consider a sequence of scales, of the same class as the Sommer scale \cite{Necco:2001xg}
\begin{equation}
r_n^2 F(r_n) = C_n.
\end{equation}
The Sommer scale $r_0$ takes $C_0 = 1.65$. Thus a step scaling scheme with scale factor $s$ can then be defined 
choosing $p_n = s^n p$ and $r_n = \frac{r_0}{s^n}$ as follows:

\begin{itemize}
\item Determine $\sigma(p_0, p_1)$ in continuum limit holding $r_0 p_0$ fixed such that $r_0^2 F(r_0) = C_0$
\item Determine $C_1 = \frac{r_0^2}{s^2} F(\frac{r_0}{s})$ in continuum limit holding $r_0$ fixed
\item Decrease $L$ by $\simeq {\frac{1}{s}}$ without fine tuning
\item Determine $\sigma(p_1, p_2)$ in continuum limit holding $r_1 p_1$ fixed such that $r_1^2 F(r_1) = C_1$
\item Determine $C_2 = \frac{r_1^2}{s^2} F(\frac{r_1}{s})$ in continuum limit holding $r_1$ fixed
\item Decrease $L$ by $\simeq {\frac{1}{s}}$ without fine tuning\\
etc...
\end{itemize}

The guideline $r < \frac{L}{3}$ should ensure finite volume safety and using the tree level improved potential \cite{Bali:2002wf} helps reduce discretization effects. Scale setting in this way will be difficult at short distances where the potential runs logarithmically however several steps should be possible before this. To investigate the volume dependence of the vertex functions themselve we performed measurements of identical operators on $\beta = 2.13$ lattices with $16^3 \times 32 \times 16$ \cite{Allton:2007hx} and $24^3 \times 64 \times 16$ \cite{Allton:2008pn} volumes. The results are shown in figure \ref{fig:voldep} (b,c,d). The volume dependence exists but is much smaller than the error from scale setting. Although extrapolations to infinite volume are possible they are probably not necessary until the error from setting the scale is reduced significantly.

\section{Results}
\begin{figure}[ht]
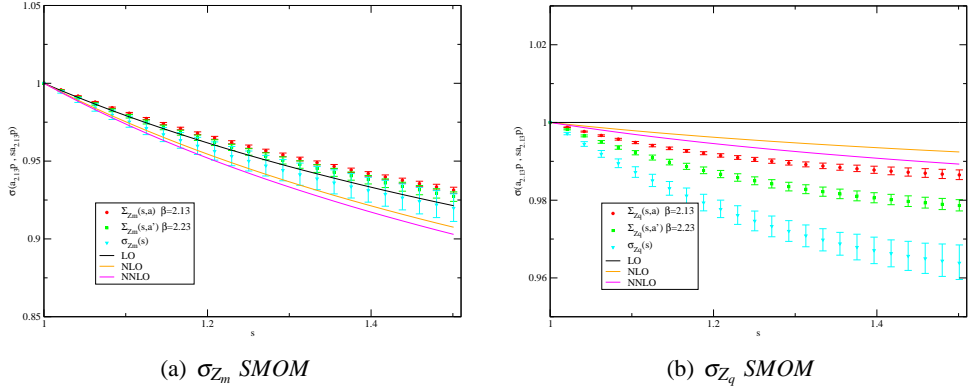

  \begin{center}
    \subfigure[$\sigma_{Z_m}$ $SMOM$]{\includegraphics[angle=0,width=0.4\textwidth]{NEZm.eps}}\,\,\,\,\,\,\,\,\,
    \subfigure[$\sigma_{Z_q}$ $SMOM$]{\includegraphics[angle=0,width=0.4\textwidth]{NEZq.eps}} 
  \end{center}
  \caption{(a) Step scaling function for $Z_m$ in the continuum limit. Details of the continuum limit are in \cite{Arthur:2010ht}. In physical units this corresponds to between $2$ and $3$ GeV. High order perturbation theory describes the running quite well. (b) The step scaling function for $Z_q$ from $2$ to $3$ GeV. Note that the continuum limit flips the direction of the running to agree, at least in sign, with perturbation theory. }
  \label{fig:Step_Zm}
\end{figure}

Full details of our ensembles have been reported elsewhere \cite{Arthur:2010ht}, here we show step scaling functions for quark mass, quark field and $B_K$ and comment on several interesting features. Figure \ref{fig:Step_Zm} (a) shows the step scaling function for the quark mass in the $SMOM$ scheme \cite{Sturm:2009kb}. The perturbative and non perturbative results are in quite good agreement. Figure \ref{fig:Step_Zm} (b) shows the quark field renormalization. At non-zero lattice spacing the perturbative running is in the opposite direction to the measured running. This highlights the dangers of applying perturbation theory at fixed lattice spacing. Only in the continuum limit is it guaranteed that perturbation theory will accurately describe the running at high energy.

\begin{figure}[ht]
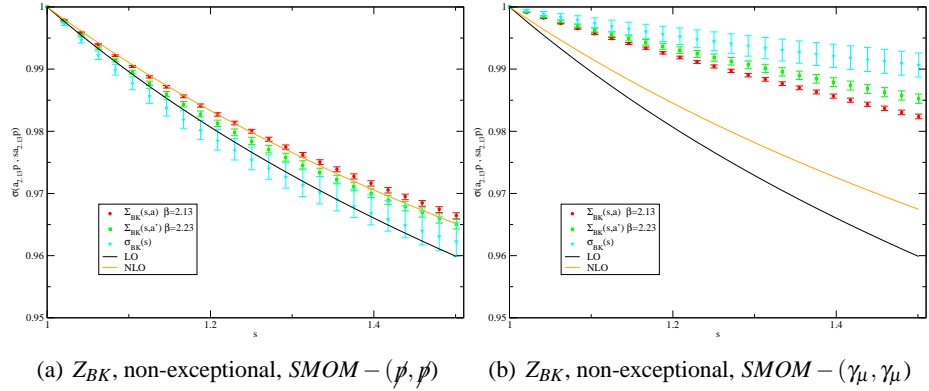

\begin{center}
\subfigure[$Z_{BK}$, non-exceptional, $SMOM-({\cancel{p},\cancel{p}}$)]{\includegraphics[angle=0,width=0.4\textwidth]{NEVVpAA_RGI.eps}}
\subfigure[$Z_{BK}$, non-exceptional, $SMOM-({\gamma_\mu,\gamma_\mu}$)]{\includegraphics[angle=0,width=0.4\textwidth]{NEVVpAA_gamma_RGI.eps}}
\end{center}
\caption{Step scaling function $B_K$ from $2$ to $3$ GeV in two schemes compared to the perturbative running.}
\label{fig:Step_BK}
\end{figure}

Figure \ref{fig:Step_BK} shows the step scaling function for $B_K$ in two schemes \cite{IWBKcontinuum}. The choice of intermediate renormalization scheme greatly affects the agreement with perturbation theory here $SMOM-(\cancel{q},\cancel{q})$ is apparently optimal. Figure \ref{fig:Step_BK}(b) shows that on our coarser lattice the perturbative and non-perturbative running agree better than in the continuum. Again this emphasises the danger of having entangled discretization and perturbative errors. 

\section{Conclusions}
Volume source, non-exceptional, twisted boundary condition renormalization is seen to give MOM scheme renormalization constants with very small statistical and systematic errors. Therefore this will be the preferred method of renormalizing matrix elements in future RBC-UKQCD measurements. Further, thanks to the excellent precision available, step scaling arbitrary operators now becomes feasible. We have shown here some promising preliminary results and are presently computing step scaling functions more accurately and for more operators. We believe this can greatly reduce the uncertainty in matching lattice calculations to perturbation theory.

\end{document}